\documentclass[showpacs,aps,preprint,graphicx,psfig]{revtex4}
\usepackage{amsmath}
\usepackage{graphicx}
\usepackage{amssymb}
\usepackage{bm}
\usepackage{color}

\begin{document}

\title{Peculiarities of clusters formation in true ternary fission of $^{252}$Cf and $^{236}$U$^*$}
\author{
A. K. Nasirov$^{1,2}$, W. von Oertzen$^{3,4}$, A. I. Muminov$^{2}$
and R. B. Tashkhodjaev $^{2}$}

\affiliation{
$^1$Joint Institute for Nuclear Research, Joliot-Curie 6, 141980 Dubna, Russia\\
$^2$Institute of Nuclear Physics, Ulugbek, 100214, Tashkent, Uzbekistan\\
$^3$Helmholtz-Zentrum Berlin, Hahn-Meitner Platz, 14109 Berlin, Germany\\
$^4$Fachbereich Physik, Freie Universit\"at, Berlin, Germany\\}

\begin{abstract}
 The existence of a new type of cluster decay called "collinear cluster tri-partition" (CCT)
 is discussed by an analysis of the landscape of the potential energy surface (PES).
 The total energy of the ternary system is
 found as a sum the binding energies of fragments and nucleus-nucleus interaction
 between them. The pre-scission state of the ternary system is assumed to be arranged as a chain
 of the three fragments along a straight line.  Minima and valleys of the PES
 are determined by variation of the proton and neutron
 distributions between them.  Pre-scission prompt emission of neutrons is assumed and
 PES is calculated for the cases of emission of 2---4 neutrons.
 The presence of the valley corresponding to the formation of the
 isotopes of Sn with masses $A$=130---136 is inherent for all PES calculated
  for CCT for spontaneous fission of $^{252}$Cf and fission induced
   by neutrons of $^{235}$U. There are local minima indicating the
   formation of Ca, Fe, Ni, Ge and Se isotopes having magic proton or/and
   neutron numbers, such as 20, 28, and 50. The analysis shows that the experimentally observed $^{68}$Ni is
   formed as the edge fragment of the ternary system connecting
   to Sn by Si and Ca isotopes at fission of $^{236}$U and $^{252}$Cf, respectively.
\end{abstract}

\date{Today}
\maketitle

\section{Introduction}

The role of the nuclear shell structure in the formation of fission
products appears in the observed asymmetric mass distributions
depending on the excitation energy of the system undergoing in fission.
In Refs.\cite{Sida,Goverd} the yields of very asymmetric Ni and even Fe isotopes
in the cold fission of $^{236}$U  analyzed as a fission channel
around magic proton number $Z=$28 with the mass number 70 are discussed.
The yield of the very asymmetric products was not observed in coincidence
with conjugate heavy product and, therefore, the authors of
the  mentioned paper did not consider the possibility of those fission
fragments as CCT products. This scenario of the yield of the
 very asymmetric products of the Ni isotopes with $A$=68 and 70
 was studied in the $^{235}$U($n_{\rm th},f$)
 reaction and at spontaneous fission of $^{252}$Cf by the FOBOS collaboration
\cite{Pyatkov2003,PyatkovEPJA45,PyatkovPHAN73,Pyatkov2011,PyatkovEPJA48}.

The observation of two and more nuclear fission products in the
fission of  $^{235}$U with thermal neutrons and in the spontaneous
fission of $^{252}$Cf has opened a new area of study in the nuclear
reactions.  This phenomenon is connected with the appearance
of cluster states in nuclear reactions and it is the manifestation of
the shell structure which is responsible for the production of
isotopes with magic numbers of neutrons and protons. When a massive
nucleus loses its stability and goes to fission, first of all
clusters are formed as future fragments having the neutron or/and
proton number nearby the magic numbers 28, 50, 82 and 126. In the
case of ternary fission one observes fragments with the charge
number 28 and 50. This kind of ternary fission is called as
the true ternary fission, which produces the fragments with
comparable masses in difference from emission of alpha-particles
or light charged nuclei $A<16$ accompanying binary fission \cite{Goenen}.

The probability of
multicluster fission of the U, Pu and Cf isotopes are less than one
percent of the corresponding cross sections of binary fission.
The cross section of CCT is comparable with one of the well-known
ternary fission with the emission of an alpha-particle  \cite{Theobald,Daniel}.
The emission of the light charged particles from the neck region on
the plane perpendicular to the fission axis
is main characteristic of the ternary fission. Although the light charged particles
can be emitted in direction close to the  momentum of the fission fragments \cite{Theobald}.
The probability of the yield of the light charged particles
 heavier than alpha-particle decreases by increasing their charge and
 mass numbers. The observed yield of
 the group of the neutron rich isotopes of Ni and Ge in coincidence
 with the heavy fragment with mass numbers $A$=136---140 is an unexpected
 phenomenon.  Therefore,
theoretical interpretation of these processes is required for a full
understanding of the mechanism.
The ternary fission fragmentation of $^{252}$Cf for all possible
third fragments using the recently proposed three-cluster
model \cite{ManimaPRC} was studied in ref.\cite{ManimaEPJA}.
The authors concluded that the theoretical relative yields imply
the emission of the $^{14}$C, $^{34,36,38}$Si,
$^{46,48}$Ar, and $^{48,50}$Ca as the most probably third
particle in the spontaneous ternary fission of $^{252}$Cf.

The results being discussed in this paper
are based on two different experiments with binary coincidences of fission
fragments and measurements of the masses and energies of the two fragments \cite{PyatkovEPJA45}.
In two other experiments \cite{PyatkovEPJA48} for the study of
spontaneous ternary fission of $^{252}$Cf, events in coincidence with neutrons are reported.
The prompt emission of neutrons from the neck region (scission neutron source) is inherent to
the spontaneous fission of the actinides \cite{Kornilov}.
Authors of Ref.\cite{PyatkovEPJA48} reported about
the two registered products of CCT in coincidence with the neutron multiplicity
emitted from the neck region of fissioning nucleus. The ``neutron belt'' was assembled
in a plane perpendicular to the symmetry axis of the spectrometer, which serves as the mean
fission axis at the same time. The third fragment of CCT was missed by registration setup.
\begin{center}
\begin{figure}
\vspace*{3.8cm}
\resizebox{1.0\columnwidth}{!}{\includegraphics{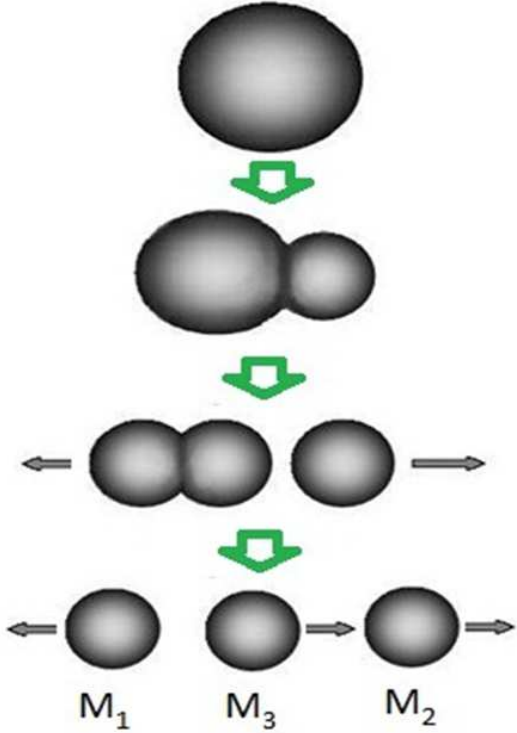}}
\vspace*{-5.60 cm} \caption{\label{stages} (Color online) The sketch
of the collinear cluster tripartition as two sequential fission
process. }
\end{figure}
\end{center}

  The relatively high yield of
the CCT-effect (more than $10^{-3}$/binary fission) is likely due to
the collective motion through very elongated (hyper-deformed)
pre-scission shapes and a large phase space covering a larger number of mass partitions with high
$Q$-values \cite{Vijayar}. The formation
of the third cluster occurs in the neck region between the main binary fragments during the
 pre-scission stage of the splitting. The fact that the formation of binary fragments is the main
 channel is seen from all PES figures as a wide and deep valley at $Z_3<2$,
 where $Z_3$ is a charge number of the middle cluster.

 The
case of alpha-cluster formation has been well studied both experimentally and theoretically.
But the CCT process is needed to be studied in detail taking into account
conditions leading to the formation of ternary system with comparable masses and
dynamics of the rupture of two necks connecting border fragments to the middle nucleus.

The aim of the calculation is to explain the possibility of the population of the Ni isotopes as
a fragment in ternary system.
The relatively large cross section of the yield of fission products with the given charge and mass numbers
 is the consequence of the population of
  the states corresponding the minima of PES.
 This is a necessary condition and the cluster can be emitted from the system
  if it is able to overcome the pre-scission barrier of the nucleus-nucleus
interactions connecting the ternary system.
Therefore, at first, it is important to analyze the PES landscape calculated for the considered
fissioning system.

\section{Outline of theoretical approach}

 The calculations are performed basing on an assumption that the third cluster has appeared in the neck region of the binary fragments due to fluctuation of
 the proton and neutron transfer between them during descent from the saddle point
 before scission point. The difference between the total energy of the ternary system
 and fissioning nucleus is used as the PES which
 shows  an effect of the nuclear shell structure on the nascent fission products.
 The PES is found as a sum the energy balance of the interacting fragments and nucleus-nucleus interaction
 between them
 \begin{eqnarray}
 \label{PES3}
\hspace*{-1.0cm}
&&U(R_1,R_2,Z_1,Z_3,A_1,A_3)=V_{1}(R_{1},Z_1,Z_3,A_1,A_3)
\nonumber\\
&&+V_2 (R_2,Z_2,Z_3,A_2,A_3) 
+V_{12}^{(Coul)} (Z_1,Z_2,R_1+R_2)\nonumber\\
&&+Q_{ggg}.
\end{eqnarray}
 Here $Q_{\rm ggg}=B_1+B_2+B_3-B_{\rm CN}$ is the balance of the fragments binding energy of
 at the ternary fission;  $V_{1}\equiv V_{13}$ and $V_{2}\equiv V_{23}$ are the nucleus-nucleus interaction
 of the middle cluster ``3'' ($A$ and $Z$ are its mass and charge numbers, respectively) with the left ``1''
 ($A_1$ and $Z_1$)  and right ``2'' ($A_2$ and $Z_2$) fragments of the ternary system;
  $V_{12}^{(Coul)}$ is the Coulomb interaction between two border fragments  ``1'' and ``2''
  which are separated by distance $R_1+R_2$, where $R_1$ and $R_2$ are the distances
  between the middle cluster and two clusters placed on the left and right sides, respectively.
  The interaction potentials $V_{1}$ and $V_{2}$ consist of the Coulomb and nuclear parts:
 \begin{eqnarray}
   &&V_{i}((R_{i},Z_i,Z_3,A_i,A_3)=V_{i}^{(Coul)}(Z_i,Z,R_i)\nonumber\\
   &&+V_{i}^{(nucl)}(Z_i,A_i,Z_3,A_3,R_i), \hspace*{0.25cm} {\rm where} \hspace*{0.25cm} i=1,2.
 \end{eqnarray}
  The nuclear
 interaction calculated by the double folding procedure with the effective
 nucleon-nucleon forces $f_{\rm eff}$ depending on nucleon density distribution  \cite{Migdal}.
 The nucleon density of fragments is used as the Fermi distribution with the parameters
 $r_0$ = 1.15--1.18 fm, $\rho_0$ = 0.17 fm$^{-3}$, $a$ = 0.54 fm. The details of the method can be
 found in the Appendix of Ref. \cite{Tashkhod}.  The Coulomb interaction is
 determined by the Wong formula \cite{Wong1973}.
\begin{figure}
\resizebox{1.0\columnwidth}{!}{\includegraphics{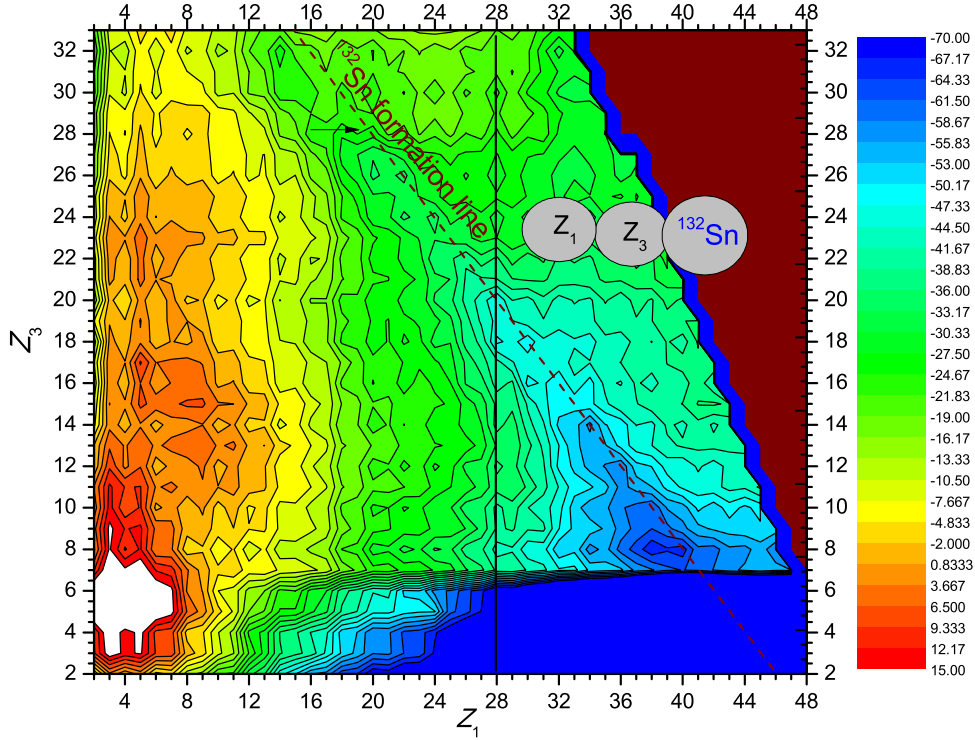}}
\vspace*{-2.5 cm} \caption{\label{FigCf2n} (Color online) Potential
energy surface calculated for the pre-scission state of the ternary
system formed after emission 2 neutrons from $^{252}$Cf as a
function of the charge numbers of middle cluster and left edge
fragment.
 }
\end{figure}

 The  nucleus-nucleus interaction potential between middle nuclei and
 left nuclei ``1'', $V_1(R)$, depends on the Coulomb interaction between ``1'' and ``2''
  nuclei and vice-versa, the last interaction affects  $V_2(R)$.
  The effect of the third fragment is important in the case of the short time between ruptures of two necks
  because the depth of the potential well will be smaller, {\it i.e.} the barrier against fission
  will be smaller. Consequently the probability of the ternary fission increases.
  The values of the binding energies of  all considered possible fragments
 as constituents of ternary system,  obtained from the table of masses
 by Audi {\it et al} \cite{Audi03}. The procedure of calculations
 of PES by Eq. \ref{PES3} has been made by following steps:
 i) We find positions $R_{\rm m1}$ and $R_{\rm m2}$ of the edge fragments
  $^{A_1}Z_1$ and $^{A_2}Z_2$ relative to the middle cluster $^{A_3}Z_3$, respectively,
   providing the minimum value of $U$ by variations of values of $R_1$ and $R_2$.
     ii)  The cluster mass number $A_3$ for each charge number $Z_3$ is
     changed from the minimum value $A_3=2Z_3$ up to $A_3^{\max}=2.6 Z_3$
     leading to a strong increase of U.
 iii) The charge number of the right edge fragment  $Z_2$  is
   found from the conservation law for the proton numbers
   $Z_2=Z_{\rm tot}-Z_3-Z_1$.
 iv) The neutron distribution between constituents of ternary system at the given charge distribution.
 In order to find the minimum of $U$ as a function of $A_1$ for the given
     mass and charge numbers $^{A_3}Z_3$ of the middle cluster and
    charge numbers $Z_1$ and $Z_2$, we vary  $A_1$  from $A_1$=2$Z_1$ up to $A_1^{\rm max}$
    corresponding to the strong increase of PES.
   The mass number of the right edge fragment  $A_2$  is found from
      $A_2=A_{\rm tot}-A_3^{\min}-A_1$. The above mentioned
    $Z_{\rm tot}$ and $A_{\rm tot}$ are the total charge and
    mass numbers of the fissioning system.

    The results of calculations  $U(R_{\rm m1},R_{\rm m2},Z_3,Z_1,A_3,A_1)$
      can be presented as a matrix with the size ($\Delta Z_3, \Delta Z_1$), where
     $\Delta Z_3$ and   $\Delta Z_1$ are the interval of variation of the
      $Z_3$ and $Z_1$, respectively.

 Probably the constituents are not in their ground state
 after formation and before their escape from the ternary system, but a procedure of calculation  by consideration mass and charge numbers as variables, changing in the wide range of  values will be very labor-consuming.

 Therefore, shell effects in the binding energies
 do not depend on their deformation.

\begin{figure}
\begin{center}
\resizebox*{1.0\columnwidth}{!}{\includegraphics{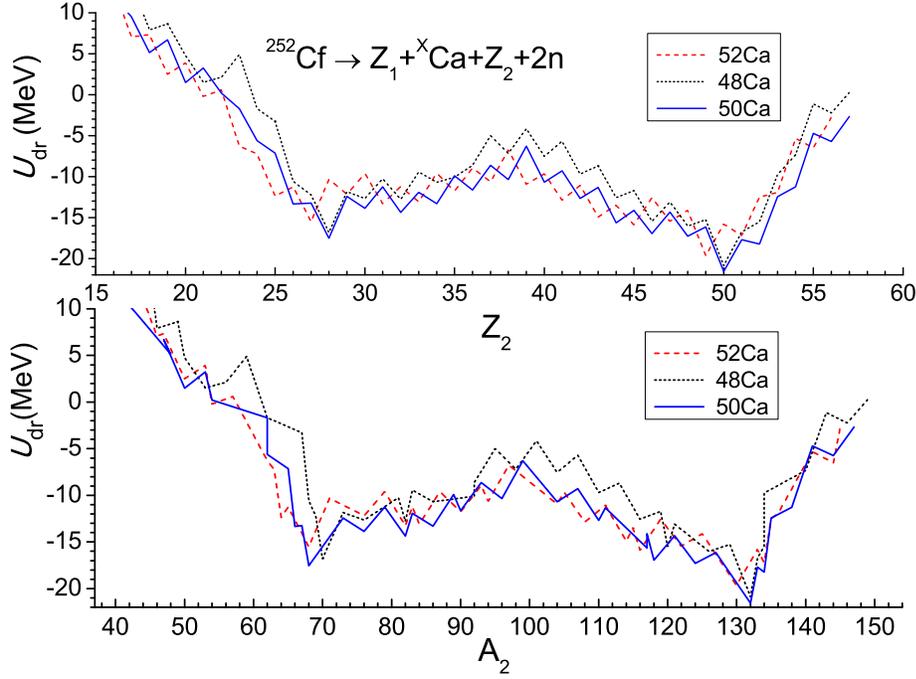}}
\vspace*{-2.5 cm} \caption{\label{UdrZ28A68} (Color online)
Comparison of the driving potentials calculated for the pre-scission
state of the collinear ternary system $Z_1+^A$Ca$+Z_2$ formed in the
spontaneous fission of $^{252}$Cf as function of $Z_2$ (upper panel)
and as function of $A_2$ (lower panel)}.
\end{center}
\end{figure}

\section{Results of calculation}

    Results of the PES for the ternary fission of $^{252}$Cf are presented by a  contour map in Fig.
\ref{FigCf2n} as a function of the charge numbers of the middle cluster $Z_3$ and one, $Z_1$, the right edge fragment.
It can be seen a valley corresponding to the formation of the cluster $^{132}$Sn for different values
of $Z_3$ and $Z_1$. This fact reflects the long tail in the mass-mass distributions of the experimentally
registered products which is parallel to the $M_1$ and $M_2$ axes (see Fig.4 in Ref.
\cite{PyatkovEPJA45} and Fig. 5a in Ref. \cite{PyatkovEPJA48}).
Those tails demonstrate the persistence of shell structure in the
double magic nucleus $^{132}$Sn
 in the formation of the fission fragments.
 The vertical line at $Z_1$=28 shows local minima corresponding to the formation of Ni isotopes
 as fragments of the ternary system. This line crosses the line on valley of formation
 $^{132}$Sn isotope at $Z_3$=20 where there is a local minimum. The probability of
 formation of the cluster configurations $^{132}$Sn+$^{50}$Ca+$^{68}$Ni after emission
 of two neutrons is large because the proton or neutron numbers of the three
 fragments are equal to the magic numbers wether 28, 50 and 82.
 The dependence of the driving potential $U_{\rm dr}$ extracted from PES
 $U(R_{\rm m1},R_{\rm m2},Z_1,Z_3,A_1,A_3)$ at $R_1=R_{\rm m1}$ and
 $R_2=R_{\rm m2}$ on the charge and mass numbers of the right edge fragment
 is shown in the upper and bottom figures of Fig. \ref{UdrZ28A68}, respectively.
 The comparison of results obtained for the mass numbers  48 (dotted curve),
  50 (solid curve) and 52 (dashed curve) of Ca being the middle cluster
  in Fig. \ref{UdrZ28A68}, demonstrates that minimal values of driving potential
  corresponds to the formation of the $^{68}$Ni isotope, which was observed with
  sufficiently large probability (see Refs. \cite{PyatkovEPJA45,PyatkovEPJA48}),
  when $^{50}$Ca is formed as a middle cluster and  $^{132}$Sn is the right edge
  fragment.
   On the contour map of the PES there are local minima
 showing the favored population of
 $^{132}$Sn+$^{38}$S+$^{82}$Ge, $^{132}$Sn+$^{36}$Si+$^{84}$Se,  $^{150}$Ba+$^{22}$O+$^{80}$Ge,
 and others.  We found that the middle cluster is more neutron rich than edge fragments.
 A much smaller energy minimum  in the PES (by 10 MeV) for the alternative configuration,
 the $^{132}$Sn+$^{72}$Ni+$^{48}$Ca channel, gives  for  this reaction a much smaller probability,
 the difference is due to the changed  Coulomb repulsion forces. This effect
 is observed in the yields observed in the experiment \cite{PyatkovEPJA45}.

\begin{figure}
\vspace*{-1.30cm}
\begin{center}
\resizebox*{1.05\columnwidth}{!}{\includegraphics{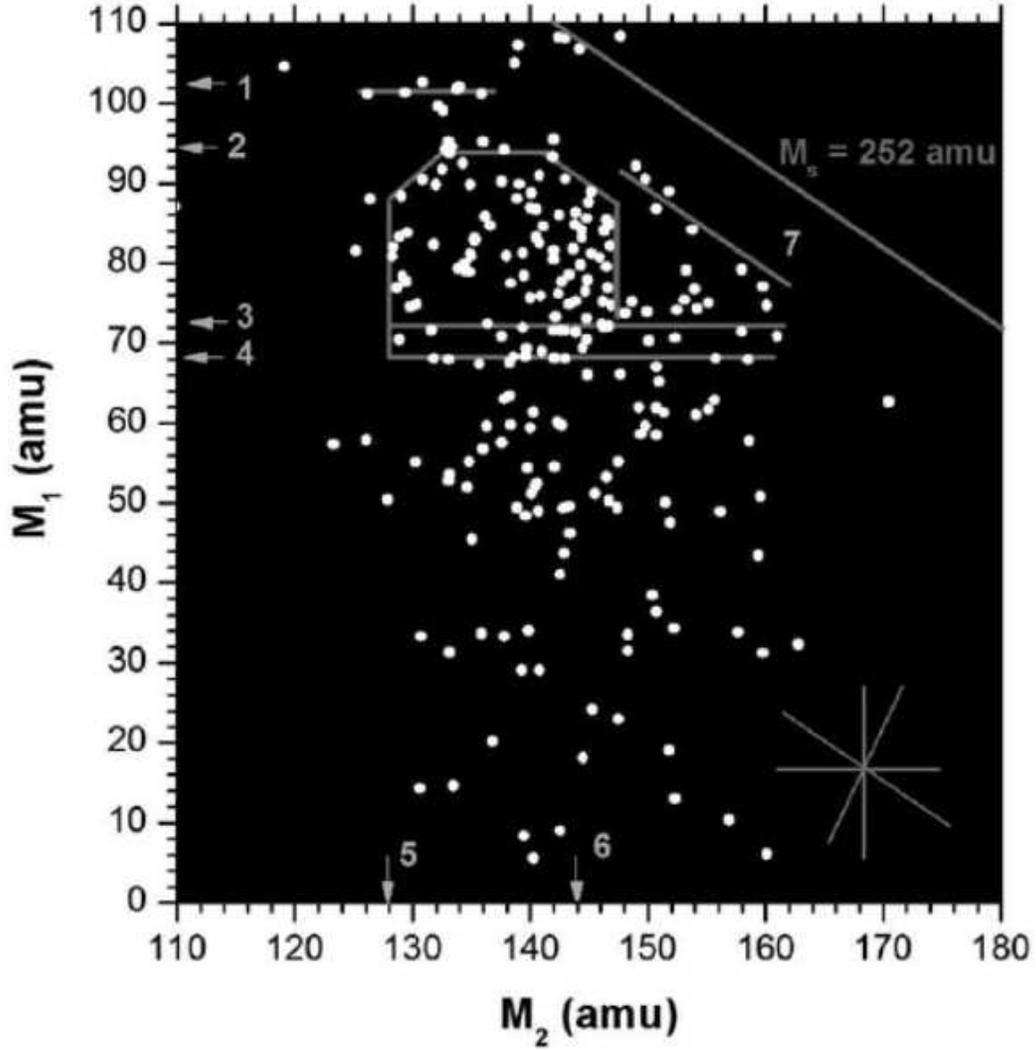}}
\vspace*{-9.2 cm} \caption{\label{ExpM1M2} The mass - mass
distribution of the fission-fragments of the spontaneous fission of
$^{252}$Cf gated by 2n emission.
 Arrows with numbers 1–6 mark the
positions of masses of magic nuclei, a line numbered 7 points
to events with the loss of a $^{14}$C nucleus.
The main intensity is with masses for the third fragments from
36–20. (Copy of Figure 10 from Ref.\cite{PyatkovEPJA48})}
\end{center}
\end{figure}

 In Fig. \ref{ExpM1M2} obtained from Ref. \cite{PyatkovEPJA48} the events of
 yields of two products in the spontaneous fission of $^{252}$Cf registered in coincidence
 are presented. Intense yield of the fission products with $A_1=$68---94,
 $A_1=$50---60, and $A_2=$128---146 registered in coincidence is observed.
 Obviously we see the sufficient influence of the shell effects in nuclear matter
 in formation of the ternary fission products.
\begin{figure}
\vspace*{-0.3cm}
\resizebox*{1.0\columnwidth}{!}{\includegraphics{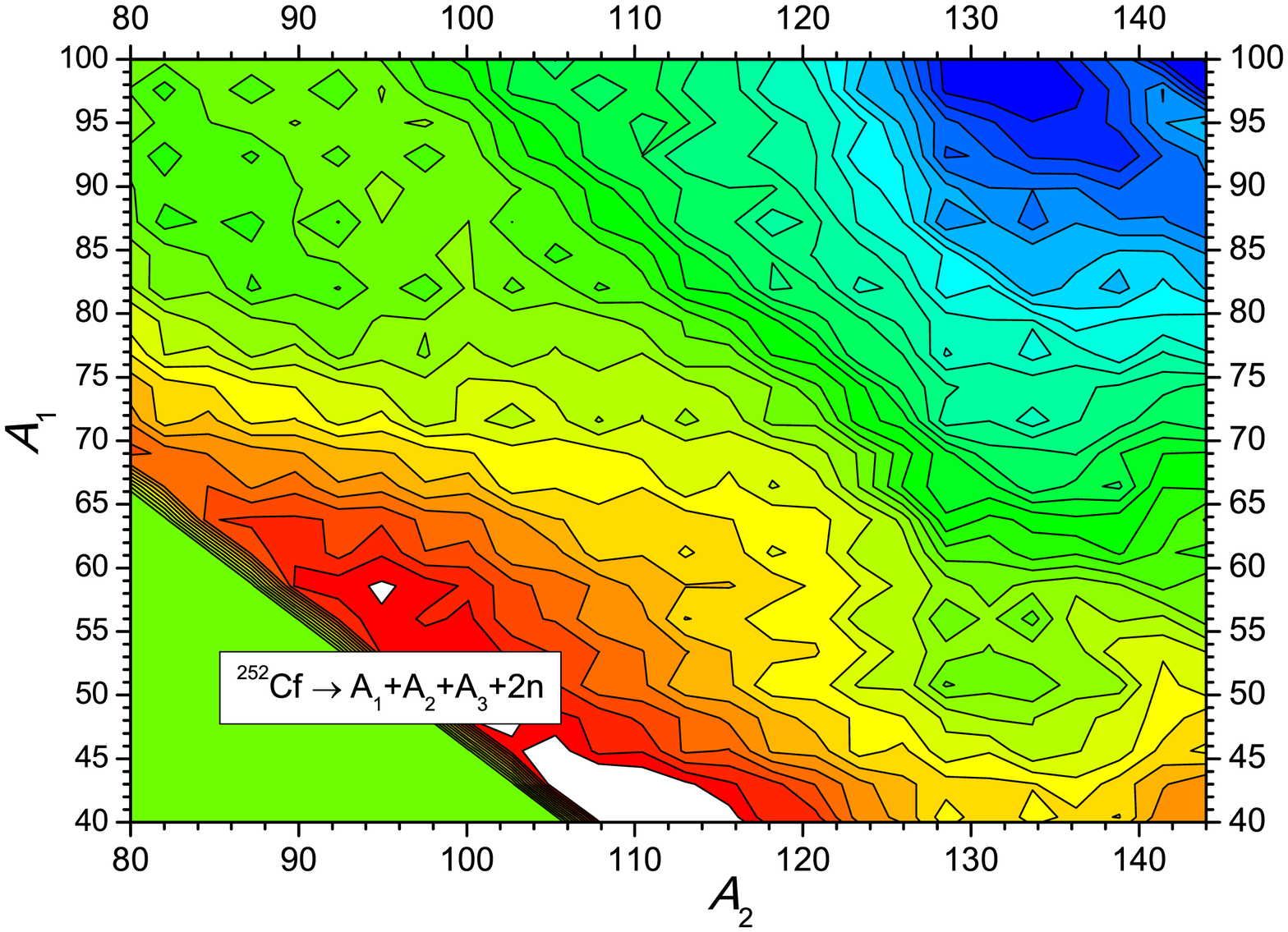}}
\vspace*{-2.80 cm} \caption{\label{ZoomCf252} (Color online)
Potential energy surface calculated for the pre-scission state of
the ternary system formed after emission 2 neutrons from $^{252}$Cf
as a function of the mass numbers of the left and left edge
fragments. }
\end{figure}

\begin{figure}
\hspace*{-0.4cm}
\resizebox*{1.0\columnwidth}{!}{\includegraphics{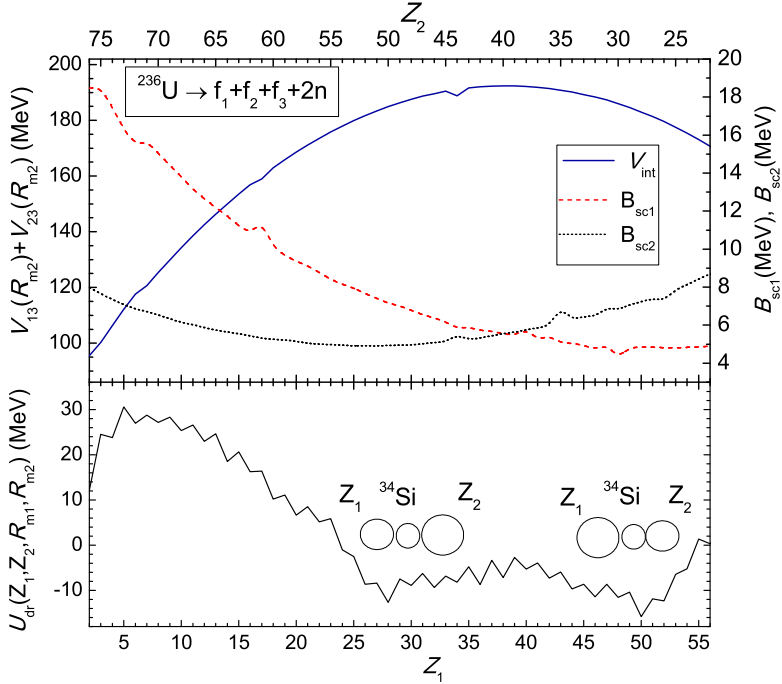}}
\vspace*{-2.4 cm} \caption{\label{U236UdrB} (Color online) The
nucleus-nucleus interaction potential (solid line, left axis),
pre-scission barriers (right axis) $B_{sc1}$ and $B_{sc2}$ for the
separation of the fragments $Z_1$ and $Z_2$ in the collinear ternary
system fragment  as a function of $Z_1$ and $Z_2$ (upper figure);
Driving potential for the  collinear ternary system as a function of
$Z_1$ when middle cluster is $^{34}$Si corresponding to the minimum
value of PES (bottom figure) for CCT in the $^{235}$U($n_{\rm
th},f$) reaction.}
\end{figure}

 The results  calculated for $^{252}$Cf  and presented in Fig. \ref{ZoomCf252}
 show the valley of minimum values of the PES in the mass number regions
 $A_1=$70---100 and $A_2=$124---144. The population of the mass distribution
 of the edge fragments  in the corresponding ranges of $A_1$ and $A_2$
 should  dominate in the ternary fission of   $^{252}$Cf .
 The fission probability from these pre-scission states depends on the
 value of the pre-scission barriers $B_{\rm sc1}$ and $B_{\rm sc2}$
 which is determined by the depth of the
 potential well.
 The splitting probability of $^{132}$Sn from the other part of system is
 determined by the depth of the potential well,  which can
 be considered as pre-scission barrier $B_{\rm sc}$. The value of  $B_{\rm sc}$
  depends on the charge distribution between fragments of the ternary system.
  For the collinear configuration of the ternary system, we have two necks
  in the connected system  and, consequently, we have two barriers  $B_{\rm sc1}$
  and $B_{\rm sc2}$ for separation of the left and right edge fragments.
  The answer to the question,  if these necks are unstable to rupture how
  the order of their scission depends
  on the relation between $B_{\rm sc1}$ and $B_{\rm sc2}$.
  The dependence of the values of $B_{\rm sc1}$ and $B_{\rm sc2}$ on the
  charge number $Z_1$, for the case that the middle cluster is $^{34}$Si at CCT of $^{236}$U,
  is shown in upper figure of Fig. \ref{U236UdrB}. The bottom part figure of Fig. \ref{U236UdrB}
  presents the driving potential of the ternary system  $Z_1+^{34}$Si$+Z_2$ being
  formed at CCT of  $^{236}$U.
  The presented results for the driving potential allow us to conclude, that
  1) formation of the
  $^{68}$Ni isotope in coincidence with $^{132}$Sn occurs with large
  probability,  because this configuration of the collinear ternary system has lower
  potential energy; the rupture of the neck connecting $^{132}$Sn to
  $^{34}$Si+$^{68}$Ni system occurs more easy than the rupture of
  the  neck connecting $^{68}$Ni to  the $^{34}$Si+$^{132}$Sn
  system. Then breaks down the $^{34}$Si+$^{68}$Ni system in the
  field of $^{132}$Sn.
  The similar situation takes place in case of CCT in the spontaneous
  fission of $^{252}$Cf, where $^{68}$Ni is formed together with
  $^{132}$Sn + $^{50}$Ca.

\section{Summary}

 A possibility of the formation of $^{68}$Ni isotope in the spontaneous
  fission of $^{252}$Cf and $^{235}$U(n$_{\rm th}$,f) reaction,
  analyzed by calculation of the potential energy surface for the collinear
  ternary system is obtained. It is found as sum of the energy balance of
  fission into three  fragments,  nucleus-nucleus interaction between neighbour fragments
  and the Coulomb interaction between the edge fragments.
  Mass and charge distributions of the collinear ternary system is
  studied by calculation and analysis of the potential energy surface
  as a function of the distances $R_1$ and $R_2$ connecting the edge
  fragments $Z_1$ and $Z_2$ to the middle cluster $Z_3$.
  This procedure has been done  changing the charge numbers $Z_1$, $Z_2$ and $Z_3$
  and corresponding mass numbers in the wide range of their values.
  As a result we determine the pre-scission state of the ternary
  system.
   The landscape  of the potential energy surface presented as function
    of fragments $Z_1$ and $Z_3$ showed well pronounced deep valley
    corresponding to the formation of $^{132}$Sn with different pair of
    fragments with $Z_1$ and $Z_3$. Minima and valleys can be seen
    on the contour map of landscape. There are local minima indicating of
   formation of Ca, Fe, Ni, Ge and Se isotopes having magic proton or/and
   neutron numbers, such as 20, 28, and 50.
  The analysis shows that the experimentally observed $^{68}$Ni is
   formed as the edge fragment of the ternary system connected
   to the formation of Sn by Si and Ca isotopes at fission of $^{236}$U and $^{252}$Cf, respectively.

\section{Acknowledgments} The authors are grateful to Drs. D.V. Kamanin and
Yu.V. Pyatkov for valuable discussions. A.K. Nasirov is grateful to
the Russian Foundation for Basic Research for the partial financial
support  of this work.


\begin{thebibliography}{99}
\bibitem{Sida}  
 J.L. Sida {\it et al.}, Nucl. Phys. A {\bf 502}, 233c (1989).
\bibitem{Goverd} 
 A.A. Goverdovsky {\it et al.},
 Phys. of Atomic Nuclei (Yadernaya Fizika) V. {\bf 58}, 1546 (1995).
\bibitem{Pyatkov2003} 
Yu. V. Pyatkov, \textit{et. al.}, Physics of Atomic Nuclei, {\bf 66},
 1631 (2003), from Yadernaya Fizika, {\bf 66}, 1679 (2003).
\bibitem{PyatkovEPJA45} 
Yu. V. Pyatkov \textit{et. al.}, Eur. Phys. J. A {\bf 45}, 29 (2010).
\bibitem{PyatkovPHAN73} 
Yu. V. Pyatkov \textit{et. al.}, Physics of Atomic Nuclei,
{\bf 73}, 1309 (2010).
\bibitem{Pyatkov2011} 
Yu. V. Pyatkov \textit{et. al.}, Int. Jour. of Mod. Phys. E {\bf
20}, 1008 (2011).
\bibitem{PyatkovEPJA48} 
Yu.V. Pyatkov {\it et al.},
    Eur. Phys. J. A {\bf 48}, 94  (2012).
    \bibitem{Goenen} 
    G. G\"oenenwein, Nucl. Phys. A {\bf 734}, 213 (2004).
 \bibitem{Theobald} 
 J.P. Theobald, P. Heeg and M. Mutterer,
 Nucl. Phys. A{\bf 502}  343c (1989).
 \bibitem{Daniel} 
 A.V. Daniel {\it et al.}, Phys. Rev. C {\bf 69}, 041305(R) (2004).
\bibitem{ManimaPRC} 
K. Manimaran, M. Balasubramaniam, Phys. Rev. C {\bf 79},
024610 (2009).
\bibitem{ManimaEPJA}  
K. Manimaran and M. Balasubramaniam, Eur. Phys. J. A {\bf 45}, 293 (2010).
\bibitem{Kornilov} 
N.V. Kornilov {\it et al.}, Nucl. Phys. A {\bf 686}, 187 (2001).
\bibitem{Vijayar}  
 K.R. Vijayaraghavan, W. von Oertzen, and M. Balasubramaniam,
 Eur. Phys. J. A {\bf 48}, 27 (2012).
\bibitem{Migdal} 
 A.B. Migdal, Theory of the Finite Fermi Systems and
Properties of Atomic Nuclei (Nauka, Moscow, 1983).
\bibitem{Tashkhod}  
R.B. Tashkhodjaev, A.K. Nasirov and W. Scheid,
Eur. Phys. J. A   {\bf 47} 136  (2011).
\bibitem{Wong1973} 
C. Y. Wong, Phys. Rev. Lett. {\bf 31}, 766 (1973).
\bibitem{Audi03} 
G. Audi, A.H. Wapstra and C. Thibault, Nucl. Phys. A {\bf 729}, 337 (2003).
\end{thebibliography}
\end{document}